\documentclass[12pt,twoside]{article}

\usepackage{a4wide}  
\usepackage{calc}
\usepackage{amsmath}
\usepackage{amsfonts,amssymb}
\usepackage[hyperindex]{hyperref} 
\usepackage{graphicx}

\newcommand{\Vol}{\mathrm{Vol}}

\newcommand{\M}{\mathcal{M}}
\newcommand{\Mink}{\mathbb{M}}
\newcommand{\C}{\mathcal{C}}
\newcommand{\Bcal}{\mathcal{B}}
\newcommand{\Scal}{\Sigma}

\newcommand{\card}{\mathbf{card}}
\newcommand{\past}{\mathbf{past}}
\newcommand{\fut}{\mathbf{future}}
\newcommand{\maxi}{\mathbf{max}}
\newcommand{\mini}{\mathbf{min}}

\newcommand{\elem}{\!\in\!}
\newcommand{\expec}[1]{\left<#1\right>}

\begin{document}


\title{\vspace{-1.5cm}
\vspace{-1.5cm}
{\normalsize \hfill IMPERIAL/TP/06/SZ/01}\\
${}$\\
${}$\\
${}$\\
\Large \textbf{Evidence for an entropy bound from fundamentally discrete gravity} 
\vspace*{0.5cm}
}
\author{{\large\em D. Rideout, S. Zohren }\\[10pt]
        {\footnotesize \rm Institute for Theoretical Physics}\\[-5pt]
        {\footnotesize \rm Imperial College}\\[-5pt]
        {\footnotesize \rm London, SW7 2AZ, UK.}\\[5pt]
        {\footnotesize \rm E-Mail: \href{mailto:d.rideout@imperial.ac.uk}{d.rideout@imperial.ac.uk},  \href{mailto:stefan.zohren@imperial.ac.uk}{stefan.zohren@imperial.ac.uk}}\vspace*{1.5cm}
 }
\date{}
\maketitle

\begin{abstract}
The various entropy bounds that exist in the literature suggest that spacetime is
fundamentally discrete, and hint at an underlying relationship between geometry
and ``information''. The foundation of this relationship is yet to be
uncovered, but should manifest itself in a theory of quantum gravity. We
present a measure for the maximal entropy of spherically
symmetric spacelike regions within the causal set approach to quantum
gravity. In terms of the proposal, a bound for the entropy contained in this region can be derived from a counting of potential ``degrees of
freedom'' associated to the Cauchy horizon of its future domain of
dependence.
For different spherically symmetric spacelike regions in Minkowski
spacetime of arbitrary dimension, we show that this proposal leads, in the
continuum approximation, to Susskind's well-known spherical entropy bound. 
\end{abstract}
\newpage

\section{Introduction}

In the history of general relativity there has been a long discussion regarding the thermodynamics of gravitational systems. One of the most famous examples is the \textit{Bekenstein-Hawking formula} for black hole entropy \cite{Bekenstein:1972tm,Bekenstein:1973ur,Bekenstein:1974ax,Hawking:1974rv}, stating that the entropy of a black hole is given by a quarter of the area of the event horizon in Planckian units
\begin{equation}\label{eq:BHbound}
S_{BH}=\frac{A}{4}.
\end{equation}
This bound is truly universal, meaning that it is independent of the characteristics of the matter system and can be derived without any knowledge of the actual microstates of the quantum statistical system.

Along this line there have been several generalizations of this entropy
bound. One is \textit{Susskind's spherical entropy bound}
\cite{Susskind:1994vu}, stating that the upper bound
for the entropy of the matter content of an arbitrary spherically symmetric
spacelike region (of finite volume) is,
\begin{equation}
S_{\mathrm{matter}}\leqslant\frac{A}{4},
\end{equation}
where $A$ is the area of the boundary of the region.\footnote{There was an earlier proposed bound by Bekenstein \cite{Bekenstein:1980jp} stating that the entropy of any weakly gravitating matter system obeys $S_{\mathrm{matter}}\leqslant 2\pi E R$, where $E$ is the energy of the matter system and $R$ the circumferential radius of the smallest sphere that contains it. If one further assumes that this bound is valid for strongly gravitating matter systems, then gravitational stability in four dimensions implies that $2E\leqslant R$ and hence $S_{\mathrm{matter}}\leqslant 2\pi E R \leqslant A/4$. One sees that in four dimensions the Bekenstein bound is stronger then the Susskind bound, however, in $d>4$ gravitational stability and the Bekenstein bound only imply that $S_{\mathrm{matter}}\leqslant(d-2) A/8$ \cite{Bousso:2000md}. Hence, the geometrical Susskind bound is arguably more fundamental.}
Even though this spacelike entropy bound cannot be generalized to arbitrary
non-spherically symmetric spacelike regions, there exists a generalization in
terms of light-sheets, namely the \textit{Bousso or covariant entropy bound}
\cite{Bousso:1999xy,Bousso:2002ju}. More precisely, let $A(\Bcal)$ be the area
of any $(d-2)$-dimensional surface $\Bcal$, then the $(d-1)$ dimensional hypersurface
$L$ is called the light-sheet of $\Bcal$ if $L$ is generated by light rays which
begin at $\Bcal$, extend orthogonally away from $\Bcal$ and have everywhere
non-negative expansion. The entropy flux through the light-sheet is then
bounded by
\begin{equation}
S(L)\leqslant\frac{A(\Bcal)}{4}.
\end{equation}
This entropy bound is also widely regarded as evidence for the \textit{holographic principle} \cite{Hooft:1993gx,Susskind:1994vu,Bousso:2002ju}, stating that the maximum number of degrees of freedom carried by $L$ is given by $A(\Bcal)/4$. 

These entropy bounds 
suggest that an underlying theory
of quantum gravity should predict the bounds from a counting of
microstates. This verification of the thermodynamic laws is an important
consistency check for any approach to quantum gravity. Further, the finiteness
of the entropy might already give some indications about the actual
microstructure of spacetime. There is a semi-classical argument that the
description of a quantum theory of gravity by a local quantum field theory in
the continuum, in the absence of a high frequency cutoff, leads to infinitely many degrees of freedom in a finite region,
and therefore to a divergence in the entropy of this region \cite{Dou:2003af,Sorkin:1997gi}.  The entropy
bounds therefore suggest that spacetime possesses a fundamental discreteness
at scales of order of 
the Planck scale. Continuum physics would then have to
emerge from this fundamental theory when making a continuum approximation at
large scales. This suggests that to obtain a theory of quantum gravity one does not have to
quantize the metric fields of the continuum geometries, but should rather find a quantum theory of the discrete structure underlying those continuum geometries
\cite{Isham:1993ji}.

In the following we introduce causal sets
\cite{Hooft:1978id,Myrheim:1978ce,Bombelli:1987aa,Sorkin:1990bh,Sorkin:1990bj} as an
approach to fundamentally discrete quantum gravity.  We discuss the importance
of causal structure within this approach, and give the correspondence between discrete
causal sets and the emergent continuum manifolds in terms of faithful
embeddings. We show how in this framework one can
derive a notion of maximum entropy from a counting of potential horizon ``degrees of
freedom''
of the fundamental theory reminiscent of former ideas in the context of black
hole entropy \cite{Sorkin:2005qx,Dou:2003af}. Using this we
formulate an entropy bound for spherically symmetric spacelike
regions within the causal set approach. We then show that in the continuum limit, for different
spherically symmetric spacelike regions in Minkowski spacetime of arbitrary
dimension, this leads to Susskind's
spherical entropy bound. \\


\textbf{Conventions:} \\

Throughout this paper we work in Planck units with
\begin{equation}
c=k_B=G_N=\hbar=1,
\end{equation}
where $c,k_B,G_N,\hbar$ denote the speed of light, Boltzmann's constant, Newton's constant and Planck's constant respectively.\footnote{It is perhaps more natural to set $8\pi G_N=1$, in which case (\ref{eq:BHbound}) becomes $S_{BH} = 2\pi A$, as is done in \cite{Sorkin:2005qx}, however in this work we keep with the more common convention $G_N=1$.} In these units the Planck length is given by 
\begin{equation}
l_p=\left(\frac{\hbar G_N}{c^3}\right)^{\frac{d-2}{2}}=1,\quad\text{for}\quad d\geqslant 3.
\end{equation}

\section{Causal sets: Fundamentally discrete gravity}

\subsection{Discreteness and causal structure as first principles}

As mentioned in the introduction, causal set theory is an approach to fundamentally discrete quantum gravity. Besides taking fundamental discreteness as a first principle, the primacy of causal structure in the continuum is the main ingredient for causal sets. 

Causal order suggests itself as a fundamental principle for quantum gravity because of the enormous amount of topological
and geometrical information which it contains \cite{valdivia}. 
It has been shown that, given merely the causal
relations of events in a spacetime manifold,
one needs only the volume measure to
recover the full geometry in the continuum \cite{Hawking:1976fe,malment}.
Furthermore, causality turned out to be a crucial ingredient for another approach to quantum gravity, namely Causal Dynamical Triangulations 
\cite{Ambjorn:2005jj}, which
has led to considerable success e.g.\ over the Euclidean approach.
 
In causal set theory, causal structure in the continuum is a reflection of an order relation among causal set ``elements''. More precisely, a \textit{causal set} is defined to be a locally finite partially
ordered set $\C=(C,\prec)$, 
namely a set $C$ together with a relation $\prec$, 
called ``precedes'', which satisfy the following axioms:
\paragraph{Transitivity:} If $x \prec y$ and $y \prec z$ then $x \prec z$, 
$\forall x,y,z \in C$;
\paragraph{Irreflexivity:} $x \not\prec x$;
\paragraph{Local Finiteness:} For any pair of fixed elements $x$ and $z$ of $C$, the set of elements lying between $x$ and $z$ is finite, $\card\{y | x \prec y \prec z \}<\infty$, where $\card X$ means the cardinality of the set $X$. \\

Of these axioms, the first two say that $\C=(C,\prec)$ is a partially ordered
set or poset. The last expression, local finiteness, is important to recover the
volume information of spacetime and is hence crucial for finding any continuum
approximation of the causal set.

In the following we will summarize some basic definitions in causal set theory
of which most are ``discrete'' versions of the analogous concepts used to
describe the causal structure of continuum spacetimes
(cf.\ App.\ \ref{app:causal}).

The \emph{past} of an element $x\in{C}$ is the subset
$\past(x)=\{y\in{C}\,|\,y\prec{x}\}$. This corresponds to $J^-(x)$ in the continuum approximation. The past of a subset of $C$ is the union of the pasts of its elements.  The \emph{future} of an element $x\in{C}$ is the subset $\fut(x)=\{y\in{C}\,|\,x\prec y\}$ which respectively corresponds to $J^+(x)$ in the continuum approximation.


An important concept for the formulation of the entropy bound from
causal set theory is that of a \emph{maximal element} in a causal set
$\C=(C,\prec)$. This is an element which has no successors, i.e. an element
$x$ for which $\nexists \, y \in C$ such that $x \prec y$. The set of all
maximal elements in $\C$ is denoted by $\maxi(\C)=\{x\in{C}\,|\nexists \, y
\in C \,\text{s.t.}\, x \prec y\}$.  In analogy, a \emph{minimal element} is
one which has no ancestors, i.e. an element $x$ for which $\nexists y \in C$
such that $y \prec x$ and the set of all minimal elements in $\C$ is denoted
by $\mini(\C)=\{x\in{C}\,|\nexists \, y \in C \,\text{s.t.}\, y \prec x\}$.

\subsection{Towards a continuum approximation}\label{sec:towards}

Having given the precise definition of a causal set one might ask the
questions:  How can one actually formulate a theory of quantum gravity using
causal sets, and how can causal sets lead to the known classical notion of
smooth Lorentzian manifolds?

The aim of causal set theory is to formulate a theory of quantum gravity using
a sum-over-histories approach, where the single histories are causal sets. In
order to recover continuum physics in a semi-classical limit,
some of the histories must be well approximated by Lorentzian manifolds. In
the following we give a definition of what it means for a causal set $\C$ to
be approximated by a Lorentzian spacetime $(\M,g)$.

Consider a strongly causal spacetime $(\M,g)$. The map $\phi : C \to \M$ from a causal set
$\C=(C,\prec)$ into a spacetime $(\M,g)$ is called a \emph{conformal embedding} 
if $x \prec y \iff \phi(x) \in J^-(\phi(y)),
\;\;\forall x, y \in C$. Consider the Alexandrov neighborhood $J^+(p)\cap J^-(q)$, for every $p,q \in \M$, which forms a basis for the manifold topology of $(\M,g)$ if $(\M,g)$ is strongly causal, which we assume throughout. The
map $\phi$ is called a \emph{faithful embedding} or \emph{sprinkling} if it has the following property: The number of elements $n$ mapped into an
Alexandrov neighborhood is equal to its spacetime volume $V$, up to
Poisson fluctuations. Thus, the probability of finding $n$ elements in
this region is given by the Poisson distribution
\begin{equation}\label{eq:poisson}
P(n) = \frac{(\rho V)^n e^{-\rho V}}{n!} \,,
\end{equation}
where $\rho=l_f^{-d}$ is the density set by the fundamental length
scale $l_f$ in $d$ dimensions. In other words, this means that in a spacetime region of volume $V$ one finds on average $N\!=\!V\rho$ elements of the casual set embedded into this region, and the fluctuations are typically of the order $\sqrt{N}$. Further, the properties of the Poisson distribution lead to local Lorentz invariance in the continuum spacetime \cite{Dowker:2003hb,Bombelli:2006nm}.
From the considerations above one assumes that the fundamental length scale will be of order of the Planck length. In Sec.\ \ref{sec:evidence} we will further predict an explicit value for the fundamental discreteness scale of spacetime which will be of the order of the Planck length. 

We say that a spacetime $(\M,g)$ \emph{approximates} a causal set $\C$ if
there exists a faithful embedding of $\C$ into $(\M,g)$.
This notion gives a correspondence between causal sets and continuum
spacetimes. In terms of this correspondence continuum causal structure arises from the microscopic order relations of the causal set elements and the continuum volume measure of a region arises from counting the number of elements comprising this region. Nevertheless, one still has to prove the uniqueness of this
correspondence (up to small fluctuations). In general the precise formulation
of such a uniqueness proof is a difficult mathematical problem, and still
remains a conjecture: the ``\emph{Hauptvermutung}'' of the
causal set approach. However, it has been proven to hold for the limiting case
$\rho\mapsto\infty$ \cite{Bombelli:1989mu} and certain progress has been made
in the generalization to large but finite $\rho$ \cite{Noldus:2003si}.

\section{An entropy bound from causal set theory}\label{sec:conjecture}

In the previous section we have given a short introduction to the causal set
approach to quantum gravity. Thereby, we gave a precise definition of the
single histories of the sum-over-histories formulation at the microscopic
scale. From a more phenomenological point of view one can raise the question
of what conclusions one can draw from this for physics in the continuum. We have already seen, for example, how spacetime volume arises from a counting
of fundamental spacetime elements. In the following we want to show how
entropy bounds could arise from a counting of potential horizon ``degrees of
freedom'' at the fundamental level, giving a microscopic origin for Susskind's
spherical entropy bound.

As already mentioned in the introduction, the spherical entropy bound states that the entropy of the matter content of a spherically symmetric spacelike region $\Scal$ (of finite volume) is bounded by a quarter of the area of the boundary of $\Scal$ in Planck units
\begin{equation}\label{eq:spericalbound}
S\leqslant \frac{A}{4},
\end{equation}
in full units $S\leqslant A k_B c^3/(4 G_N \hbar)$, where $A=\Vol (\Bcal(\Scal))$ is the area of the boundary of this region. 

Surprisingly, the maximum entropy in \eqref{eq:spericalbound} can be determined
without any knowledge of the microscopic properties of the thermodynamic
system. A theory of quantum gravity however should be able
to deduce \eqref{eq:spericalbound} purely from a counting of the fundamental
degrees of freedom at the microscopic level.
(One may wonder in what manner a
counting of degrees of freedom measures the entropy of a system.  In a
discrete context ``degrees of freedom'' are generally finite, and a state
counting can be expected to yield something proportional to the exponential of
the number of degrees of freedom.  Thus measuring the entropy as the logarithm of
the number of states can be seen to be equivalent to counting the number of
degrees of freedom of the system.)
In the following we want to give a notion of those fundamental degrees of
freedom in the context of causal set theory leading to the formulation of an
entropy bound within this approach.

\begin{figure}
\begin{center}
\includegraphics[width=3in]{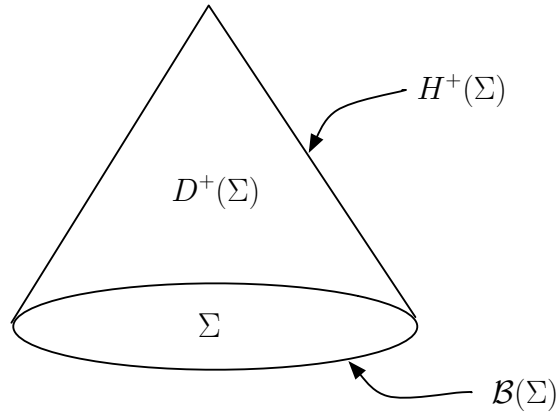}
\caption{Illustration of a spacelike hypersurface $\Scal$, its boundary $\Bcal(\Scal)$, its future domain of dependence $D^+(\Scal)$ and its future Cauchy horizon $H^+(\Scal)$.}\label{fig:domain}
\end{center}
\end{figure}

Consider a spherically symmetric spacelike region of spacetime, potentially
containing some matter system.  We assume that the ``back reaction'' of the matter
content upon the spacetime geometry can be neglected, so that different states
of the matter system lead to the same spherically symmetric spacetime geometry.
The entropy of that system
must eventually ``flow out'' of the region by passing over the
boundary of its future domain of dependence.  But because spacetime is
fundamentally discrete, the amount of such entropy flux is bounded above
by the number of discrete elements comprising this boundary.  This is a
fundamental limit imposed by discreteness on the amount of information flux
which can emerge from the region, and therefore on the amount of entropy which
it can contain.\footnote{Those familiar with causal sets might worry that we
are neglecting non-local effects when making this claim.  However at some
scale above the discreteness scale we must recover an effectively local
dynamics, thus justifying our claim (perhaps up to some locality scale
correction factor).}
Thus we have argued that an entropy bound arises from causal set quantum gravity.

\vspace{2mm}
\noindent\textbf{Proposal}\hspace{2mm}
\textit{Consider a spherically symmetric spacelike hypersurface $\Scal$ of
finite volume in a strongly causal spacetime $\M$ of dimension $d\geqslant
3$. Denote the future domain of dependence of this hypersurface by
$D^+(\Scal)$ (c.f.\ App.\ \ref{app:causal} and Fig.\ \ref{fig:domain}). Let
$\C=(C,\prec)$ be a causal set which can be faithfully embedded into
$D^+(\Scal)$. Then the maximum entropy contained in $\Scal$ is given by the number of
maximal elements of $\C$,
\begin{equation}
S_{\max}=\card\{\maxi(\C)\}.
\label{eq:proposal}
\end{equation}
}
\vspace{2mm}
\noindent\textbf{Claim}\hspace{2mm}
\textit{This proposal leads to Susskind's entropy bound in the continuum approximation,
\begin{equation}
S_{\max}= \frac{A}{4},
\end{equation}
where $A=\Vol (\Bcal(\Scal))$ is the area of the boundary of this region $\Scal$, if the fundamental discreteness scale is fixed at a dimension dependent value to be calculated.
}

\section{Evidence for the claim}\label{sec:evidence}

In this section we provide analytical and numerical evidence for the claim that the entropy bound \eqref{eq:proposal} leads to Susskind's bound in the continuum.  In all discussed
examples we consider $(d-1)$-dimensional spherically symmetric spacelike
hypersurfaces $\Scal$ in $d$-dimensional Minkowski space $\Mink^d$, and
calculate the number of maximal elements in its domain of dependence
$D^+(\Scal)$. Since $D^+(\Scal)$ is supposed to arise as a continuum
approximation of a causal set $\C$, we know from causal set kinematics
that the elements of $\C$ are faithfully embedded into $D^+(\Scal)$
according to the Poisson distribution
\eqref{eq:poisson}. Hence the expected number of maximal elements in
$D^+(\Scal)$ is given by
\begin{equation}\label{eq:defentint}
\expec{n}=\rho \int_{D^+(\Scal)} dx^d \exp\left\{-\rho \,\Vol\left(J^+(x)\bigcap D^+(\Scal)\right)\right\},
\end{equation}
where again $\rho$ is the fundamental density of spacetime. On the right hand
side of \eqref{eq:defentint} one integrates over all points $x\in
D^+(\Scal)$, where every point is first weighted by the probability of finding
an element of $\C$ embedded at this point and further weighted by the
probability of not finding any other element of $\C$ embedded in
$J^+(x)\bigcap D^+(\Scal)$.
Note that because of the fundamentally random nature of the discrete-continuum
correspondence in causal sets we calculate the expected number of maximal
elements in a sprinkling, even though the proposal (\ref{eq:proposal}) is phrased in terms of a
fixed causal set.

\subsection{The ball in Minkowski spacetime}

In this section we want to calculate the expected number of maximal elements
$\expec{n}$ in $D^+(\Scal)$, where $\Scal$ is chosen to be a
$(d-1)$-dimensional ball $S_{d-1}(R)$ with radius $R$ in Minkowski spacetime
$\Mink^d$. Due to the spherical symmetry of the problem it is useful to
introduce spherical coordinates $x=(t,r,\theta_1,...,\theta_{d-2})$, where we
choose the origin to be the futuremost event of $D^+(\Scal)$. The volume element
$\Vol\left(J^+(x)\bigcap D^+(\Scal)\right)$ is equal to the volume of the
Alexandrov neighborhood of $x$ and $0$, $\Vol_d(\tau)\!\equiv\!\Vol(J^+(x)\bigcap
J^-(0))$, where we denote proper time by $\tau=\sqrt{t^2-r^2}$. Using the
result for $\Vol_d(\tau)$ as calculated in App. \ref{app:volume} and
integrating out the spherical symmetry in \eqref{eq:defentint} one obtains a
general expression for the expected number of maximal elements in
$D^+(S_{d-1}(R))$,
\begin{equation}\label{eq:minkd}
\expec{n}=\rho\, \frac{(d-1)\pi^{\frac{d-1}{2}}}{\Gamma(\frac{d+1}{2})} \int_0^R dt \int_0^t dr\, r^{d-2}e^{-\rho D_d(t^2-r^2)^\frac{d}{2}},
\end{equation}
where the dimension dependent constant $D_d$ is defined in App.\ \ref{app:volume}.

In the following we evaluate \eqref{eq:minkd} for various dimensions by analytical and numerical methods.

\subsubsection{2+1 dimensions}\label{sec:ball2+1}

In $d=3$ dimensions one can explicitly evaluate \eqref{eq:minkd}. It is useful
to express the result in terms of the expected total number of spacetime elements
faithfully embedded into $D^+(S_{2}(R))$, $N=\rho V$, where $V=
\frac{\pi}{3}R^3$ is the volume of the domain of dependence
$D^+(S_{2}(R))$. 
The expected number of maximal elements is
\begin{equation}\label{eq:nmax2+1}
\expec{n}=8\left(e^{-\frac{N}{4}}-1\right)-2 N E_{\frac{1}{3}}\left(\frac{N}{4}\right)+4\Gamma\left(\frac{2}{3}\right)\sqrt[3]{2 N},
\end{equation}
where $E_n(x)$ is the exponential integral defined by
\begin{equation}
E_n(x)=\int_1^\infty t^{-n} e^{-xt}dt.
\end{equation}

We also measured the expected number of maximal elements numerically by
``sprinkling'' a causal set into a 2+1-dimensional $2 \times 2 \times 1$ square box, which
contains $D^+(S_{2}(1))$.  By sprinkling we mean simply selecting $N_c$
elements at random with uniform distribution within the box, and computing the
causal relation between each pair from the Minkowski metric.  For each of 100 trials $i=1\ldots 100$, we deduce
the set of elements which fall within $D^+(S_{2}(1))$, compute its cardinality
$N_i$, and count the number of such elements $n_i$ which are maximal within that
region.  From these we compute the sample mean and its error, and 
repeat this
computation for a range of values of $N_c$.  These computations were greatly
facilitated by utilizing causal set and Monte-Carlo toolkits
within the Cactus computational framework \cite{cactus}.

\begin{figure}[t]
\begin{center}
\includegraphics[width=6in]{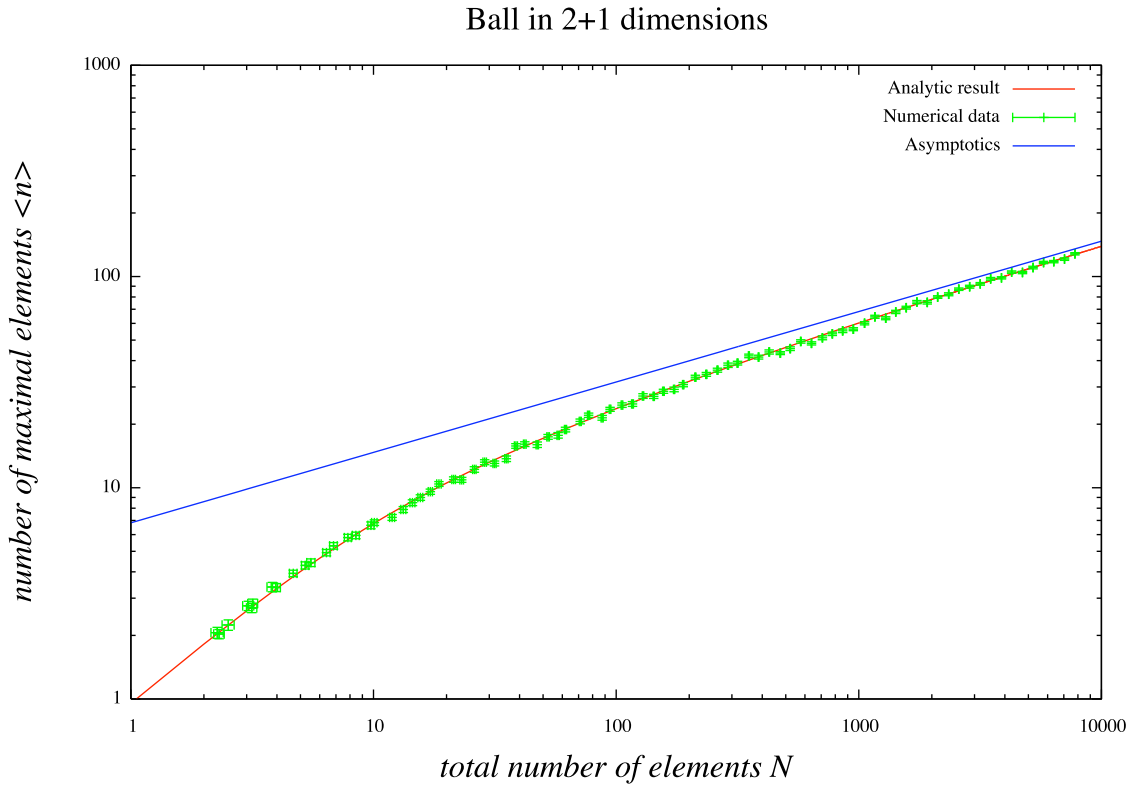}
\caption{Shown is the expected number of maximal elements $\expec{n}$ as a
  function of the total 
  number of elements in the domain of dependence of the
  unit disk in 2+1 dimensional Minkowski spacetime, on a logarithmic
  scale. Besides the analytical result and its asymptotics, data points
  with error bars from Monte-Carlo simulations are also shown.}\label{fig:disc2+1}
\end{center}
\end{figure}

The plot of the expected number of maximal elements $\expec{n}$ as a function
of $N$ for the unit disk is shown in Fig.\ \ref{fig:disc2+1},
on a logarithmic scale. 
The
agreement of analytical and numerical results justifies the 
the numerical methods. 

For a large number of elements $N\!\rightarrow\!\infty$ we can use the asymptotic expansion of the exponential integrals
\begin{equation}\label{eq:Enasymp}
E_n(x)\propto \frac{e^{-x}}{x}\left(1-\frac{n}{x}+...\right)\quad\text{for}\quad |x|\rightarrow\infty
\end{equation}
yielding
\begin{equation}
\expec{n}=4\,\Gamma\left(\frac{2}{3}\right)\sqrt[3]{2 N}+\ldots\quad\text{for}\quad N\rightarrow \infty,
\end{equation}
where $+\ldots$ are lower order terms in $N$.
The asymptotics are also displayed in Fig.\ \ref{fig:disc2+1} together with the full expression for the expected number of maximal elements.
In terms of the density $\rho$ and the radius $R$ of $S_2(R)$ this result reads
\begin{equation}
\expec{n}=\frac{16\sqrt[3]{2\pi}}{3^{5/6}\Gamma\left(\frac{1}{3}\right)}\, \rho^{\frac{1}{3}} \frac{2\pi R}{4},
\end{equation}
up to lower order corrections.
It is important
to see that $\expec{n}\!\propto\!A/4$, where $A=2\pi R$ is the length of the
boundary of $S_{2}(R)$. This is highly non-trivial, as one can see by looking
at a
snapshot of a numerical simulation (Fig.\ \ref{fig:disc2+1}). There one observes that
the maximal elements do not align along the one-dimensional boundary
$\Bcal(S_2(R))$.  Instead they are distributed along a hyperbola close to the
two-dimensional Cauchy horizon $H^+(S_2(R))$, with a density of maximal
elements which \emph{decreases} with distance from the center. Hence the fact that the
expected number of maximal elements is indeed proportional to the length $A$
of the one-dimensional boundary $\Bcal(S_2(R))$ for large $A$ already gives very
non-trivial evidence for the proposed entropy bound.

\begin{figure}[t]
\begin{center}
\includegraphics[width=4.5in]{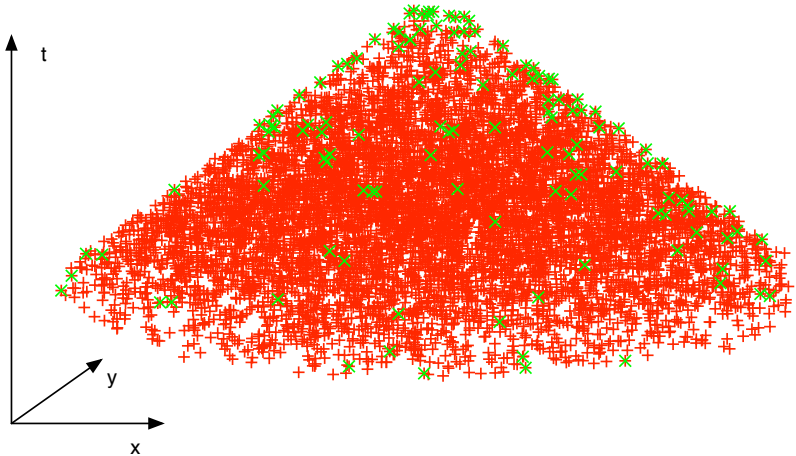}
\caption{A snapshot from a simulation showing $N=7806$ spacetime elements
forming the domain of dependence of the two-dimensional ball in 2+1
dimensional Minkowski spacetime (illustrated in red) and $n=126$
maximal elements therein (illustrated in green).}\label{}
\end{center}
\end{figure}

To have the precise confirmation of the $\mathcal{O}(1)$ constant in
$\expec{n}\!\propto\!A/4$ one has to choose the fundamental length scale
$l_f=\rho^{-1/3}$ to be $l_f=16\sqrt[3]{2\pi}/(3^{5/6}\Gamma(1/3))\approx4.41$. This gives support to the belief 
that the fundamental discreteness scale
of spacetime is of the order of the Planck length. Further, it gives a first
determination for the coefficient in 2+1 dimensions.

Hence, using this value for the fundamental discreteness scale we have confirmed our claim, namely in the continuum approximation we have
\begin{eqnarray}\label{eq:res2+1}
S_{\max}=\expec{n}=\frac{2\pi R}{4}
\end{eqnarray}
or $S_{\max}=k_B 2\pi R/(4\sqrt{\hbar G_N/c^3})$ in full units, when we set
$l_f=16\sqrt[3]{2\pi}/(3^{5/6}\Gamma(1/3))$.  
One now sees
that the assumption of large but finite $N$ used to obtain the asymptotic
behavior is equivalent to saying that $A$ is much larger than the Planck
length. However, one can see that even for relatively small values of the
length $A$, such as $10^3$ in Planck units, the approximation of $\expec{n}$
by its asymptotic expansion is already very accurate. For even smaller values
of the length scale the Planck corrections become significant. However, as
one can see in Fig.\ \ref{fig:disc2+1}, the corrections always decrease the
expected number of maximal elements, such that $\expec{n}$ never exceeds the
bound \eqref{eq:res2+1}.

\subsubsection{3+1 dimensions}

Clearly from a physical point of view the evaluation of \eqref{eq:minkd} in 3+1 dimensions is the most interesting case. For $d=4$ one can write \eqref{eq:minkd} as follows,
\begin{equation}\label{eq:3+1firststep}
\expec{n}=2\pi\rho\int_0^R dt\int_0^{t^2}dz \sqrt{t^2-z} e^{-\rho D_4 z^2}
\end{equation}
One can perform the first integration in \eqref{eq:3+1firststep} by using the following integral relation \cite{gradshteyn}
\begin{eqnarray}
&&\int_0^u x^{\nu-1}(u-x)^{\mu-1} e^{\beta x^n} dx = B(\mu,\nu) u^{\mu+\nu-1}\times\nonumber\\
&\times& {}_nF_n\left(\frac{\nu}{n},\frac{\nu+1}{n},...,\frac{\nu+n-1}{n};\frac{\mu+\nu}{n},\frac{\mu+\nu+1}{n},...,\frac{\mu+\nu+n-1}{n};\beta u^n\right), \label{eq:integrationrel1}
\end{eqnarray}
for $\Re(\mu)>0$, $\Re(\nu)>0$ and $n=2,3,...$, where $B(\mu,\nu)$ denotes Euler's beta function and ${}_pF_q(a_1,...,a_p;b_1,...,b_q;z)$ is the generalized hypergeometric function defined through
\begin{equation}
{}_pF_q(a_1,...,a_p;b_1,...,b_q;z)=\sum_{k=0}^\infty\frac{z^k}{k!}\frac{\prod_{i=1}^p (a_i)_k}{\prod_{j=1}^q (b_j)_k},
\end{equation}
and $(a)_n=\Gamma(a+n)/\Gamma(a)$ are the usual Pochhammer polynomials. The second integration, namely the one of the generalized hypergeometric function, can be obtained by use of the following relation \cite{gradshteyn}
\begin{equation} \label{eq:integrationrel2}
\int z^{\alpha-1} \,{}_pF_q(a_1,...,a_p;b_1,...,b_q;z) dz= \frac{z^\alpha}{\alpha}\, 
{}_{p+1}F_{q+1}(a_1,...,a_p,\alpha;b_1,...,b_q,\alpha+1;z).
\end{equation}
Expressed in terms of the number of causal set elements sprinkled into $D^+(S_{3}(R))$, $N=\rho V$, where the volume of $D^+(S_{3}(R))$ is given by $V= \frac{\pi}{3}R^4$, the final result reads
\begin{equation}\label{eq:nofN3+1}
\expec{n}=N\, {}_3F_3\left(\frac{1}{2},1,1;\frac{5}{4},\frac{7}{4},2;-\frac{1}{8}N \right).
\end{equation}

\begin{figure}[t]
\begin{center}
\includegraphics[width=6in]{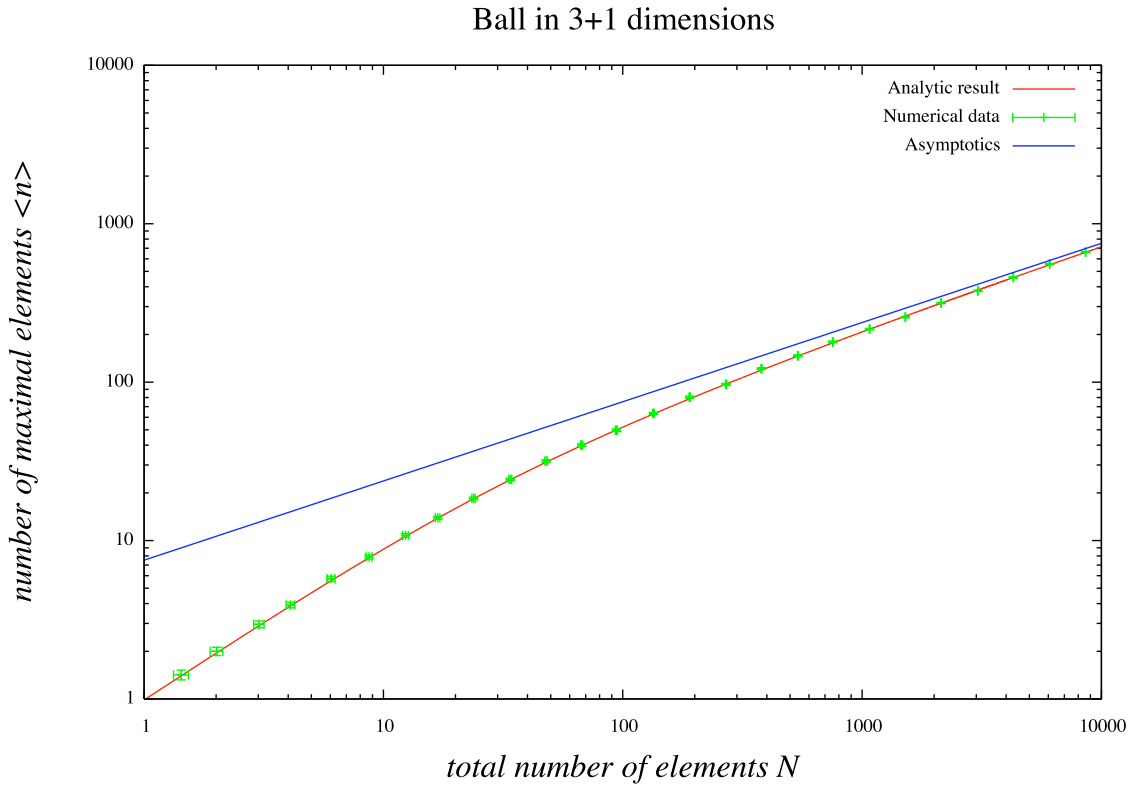}
\caption{Shown is the expected number of maximal elements $\expec{n}$ as a
  function of the total number of elements in the domain of dependence of the
  3-dimensional unit ball in 3+1 dimensional Minkowski spacetime on a
  logarithmic scale. The plot shows the analytical result, its asymptotics,
  and numerical results from Monte-Carlo simulation.}\label{fig:disc3+1}
\end{center}
\end{figure}

For a large number of elements $N$ we can use the asymptotic expansion of the generalized hypergeometric functions (cf.\ App.\ \ref{app:asymp}), yielding the asymptotic expression for the expected number of maximal elements,
\begin{equation}
\expec{n}=3\sqrt{2\pi} N^{\frac{1}{2}}+...\quad\text{for }N\rightarrow \infty.
\end{equation}
In terms of the fundamental density of spacetime $\rho$ and the radius $R$ of $S_3(R)$ this result translates into
\begin{equation}
\expec{n}=\sqrt{6}\rho^{\frac{1}{2}}\frac{4\pi R^2}{4}.
\end{equation}
As in the previous case the number of maximal elements follows the right scaling according to the entropy bound, namely $\expec{n}\!\propto\!A/4$, where the area of the boundary of 
$S_{3}(R)$ is given by $A=4\pi R^2$. The factor of proportionality is a $\mathcal{O}(1)$ constant as in the previous case, supporting the assumption that the fundamental length scale $l_f$ is proportional to the Planck length $l_p$. More precisely, to have an exact agreement with the Susskind bound the fundamental length scale in four dimensions will be given by $l_f\!=\!\sqrt[4]{6}$ in Planck units.
Using this value for the fundamental length scale the asymptotic expansion for the expected number of maximal elements reads
\begin{eqnarray}
S_{\max}=\expec{n}=\frac{4\pi R^2}{4}
\end{eqnarray}
or $S_{\max}=4\pi R^2 k_B c^3/(4 \hbar G)$ in full units. At first sight one
might feel uncomfortable in absorbing the order one constant into the
fundamental length scale to exactly confirm the Susskind bound. 
However, as discussed in the previous case, the scaling
$\expec{n}\!\propto\!A/4$ is nontrivial and already serves as a
confirmation of the bound. Taking the phenomenological law $S_{max}=A/4$
(spherical entropy bound) as ``data'' gives the
fundamental length scale in four dimensions to be
\begin{equation}
l_f=\sqrt[4]{6} \approx 1.57\quad\text{(in four dimensions)}.
\end{equation}
In comparison to the previous case of 2+1 dimensions one observes that the
fundamental discreteness scale $l_f$ depends on the dimension. In
Sec.\ \ref{sec:ballhigherd} we will derive an expression for the factor for
arbitrary (even) dimensions, showing that this factor tends exactly to one as
$d\!\rightarrow\!\infty$. It is important to check that $l_f$ is universal for
all cases with the same spacetime dimension. In Sec. \ref{sec:hyp} we will
provide evidence that this is indeed the case.

At this point it is interesting to note that a similar method of fixing the
fundamental discreteness scale to obtain the right factor of proportionality
in the entropy bound is followed in loop quantum gravity in the context
of black hole entropy (cf.\ \cite{Meissner:2004ju}). There one fixes the
Immirzi parameter which can be regarded as a measure for the discreteness
scale (through its relation to the lowest eigenvalue of the area operator) to
obtain the right factor of a quarter in the black hole entropy. However, since
there are several ambiguities in the relation between the Immirzi parameter
and the fundamental discreteness scale in the sense one uses it in causal set
theory, it is hard to compare the numerical values coming from the two derivations.

\subsubsection{4+1 dimensions}

In $d=5$ dimensions one can also evaluate the integral in \eqref{eq:minkd} in a
similar way to the calculation in 2+1 dimensions. As in the previous cases the
result is expressed in terms of the number of causal set elements
 sprinkled into $D^+(S_{4}(R))$, $N=\rho V$, with the volume of $D^+(S_{4}(R))$ given by $V= \frac{\pi^2}{10}R^5$. The final result reads
\begin{eqnarray}
\expec{n}&=&\frac{4}{3}\left(32\left(1-e^{-\frac{N}{16}}\right)+3 N E_{\frac{1}{5}}\left(\frac{N}{16}\right)-N E_{\frac{3}{5}}\left(\frac{N}{16}\right)+\right.\nonumber\\
&&\left. +\, 5 (2N)^{\frac{3}{5}}\Gamma\left(\frac{7}{5}\right)-30 (2N)^{\frac{1}{5}}\Gamma\left(\frac{9}{5}\right)\right).\label{eq:max4+1}
\end{eqnarray}

\begin{figure}[t]
\begin{center}
\includegraphics[width=6in]{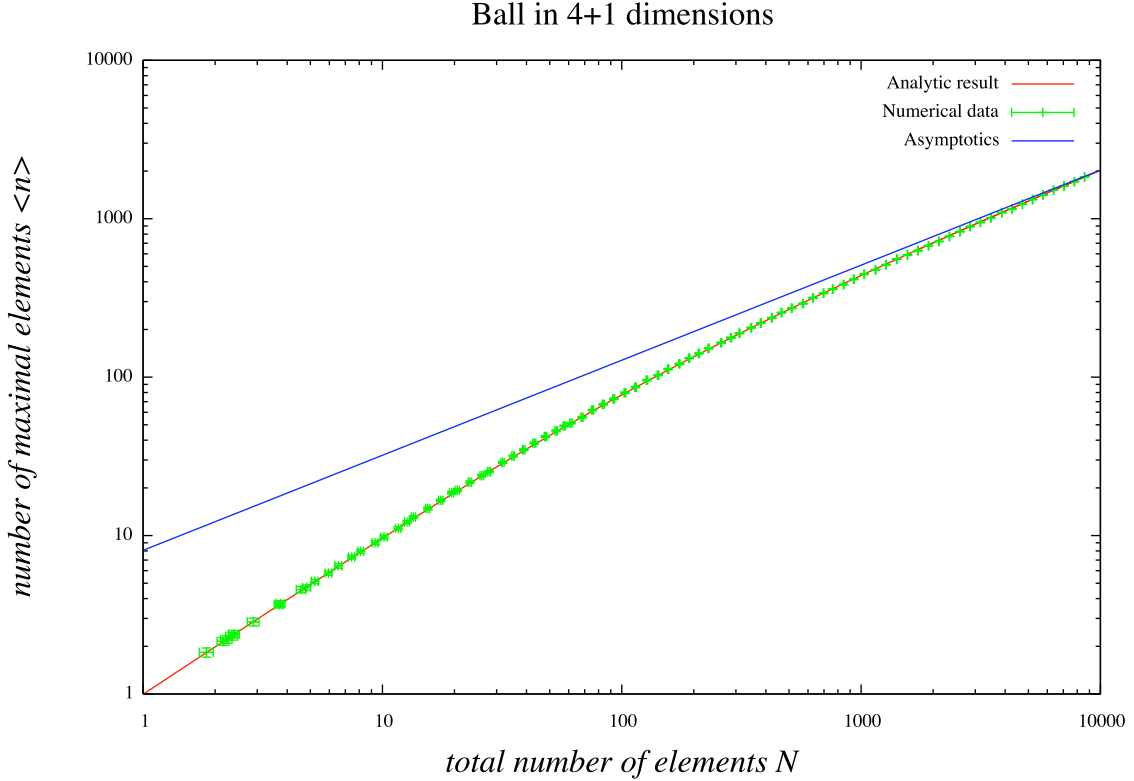}
\caption{Shown is the plot of the number of maximal elements $\expec{n}$ as a function of the total number of elements in the domain of dependence of the 4-dimensional ball in 4+1 dimensional Minkowski spacetime on a logarithmic scale, its asymptotics, and numerical results.}\label{fig:disc4+1}
\end{center}
\end{figure}

The plot of $\expec{n}$ as a function of $N$ is shown in Fig.\ \ref{fig:disc4+1}
on a logarithmic scale as well as the numerical results obtained from
Monte-Carlo simulation and the asymptotic behavior. As in the previous cases
there is agreement between analytical and numerical results.

For  large $N$ one can expand \eqref{eq:max4+1} using the asymptotic expansion for the exponential integral \eqref{eq:Enasymp}, yielding
\begin{equation}
\expec{n}=\frac{20}{3}\Gamma\left(\frac{7}{5}\right)(2 N)^\frac{3}{5}+...\quad\text{for}\quad N\rightarrow\infty.
\end{equation}
Re-expressed in terms of the density $\rho$ and the radius $R$ of $S_4(R)$ this result reads
\begin{equation}\label{eq:max4+1exp}
\expec{n}=\frac{32\pi^{1/5}}{5^{8/5}\Gamma\left(\frac{8}{5}\right)}\sqrt{\frac{2}{5+\sqrt{5}}}\, \rho^{\frac{3}{5}} \frac{2\pi^2 R^3}{4}.
\end{equation}
As for the lower dimensional cases this result gives the correct behavior
$\expec{n}\!\propto\!A/4$, where $A=2\pi^2 R^3$ is the volume of the boundary of
$S_{4}(R)$, providing further evidence for the claim. Further,
we can use the coefficient in \eqref{eq:max4+1exp} to fix the fundamental
length scale in 4+1 dimensions $l_f=\rho^{-1/5}$ to be $l_f =2^{11/6}\pi^{1/15}/(5^{8/15}(5+\sqrt{5})^{1/6}\Gamma(8/5)^{1/3})\approx 1.22$ in Planck units. 
Using this fundamental length scale the number of maximal elements \eqref{eq:max4+1exp} reads
\begin{eqnarray}
S_{\max}=\expec{n}= \frac{2\pi^2 R^3}{4},
\end{eqnarray}
or in full units $S_{\max}=2\pi^2 R^3 k_B/(4 (\hbar G_N/c^3)^{3/2})$, which confirms our claim.

\subsubsection{Generalizations to higher dimensions}\label{sec:ballhigherd}

In the previous sections we have seen that causal set theory can provide a fundamental explanation for Susskind's entropy bound
for the spacelike hypersurface $S_{d-1}(R)$ in 2+1, 3+1 and 4+1-dimensional Minkowski spacetime. In this section we want to generalize these
results to arbitrary dimensions. For the case of odd spacetime dimension, it turns out
that one can easily calculate the expected number of maximal elements, but one cannot write the results in a closed
form. However, for even dimensions one can find a closed expression.

As in the previous cases we will express the result of the expected number of maximal elements in terms of $N\!=\!\rho V$, where $V$ is the volume of the domain of dependence of $S_{d-1}(R)$, given by
\begin{equation}
V\equiv\Vol\left(D^+(S_{d-1}(R))\right)= \frac{\pi^{\frac{d-1}{2}}}{d\,\Gamma(\frac{d+1}{2})} R^d.
\end{equation}
Using the integration relations \eqref{eq:integrationrel1} and \eqref{eq:integrationrel2} one can integrate \eqref{eq:minkd} for even dimensions, yielding
\begin{equation}\label{eq:maxgenerald}
\expec{n}=N\,{}_{\frac{d}{2}+1}F_{\frac{d}{2}+1}\left( \frac{2}{d},\frac{4}{d},...,\frac{d}{d},1; 
\frac{d+1}{d},\frac{d+3}{d},...,\frac{2d-1}{d},2;-2^{1-d}N
\right).
\end{equation}
Note that this result is also valid for the case of $d=2$ dimensions, however
it is not related to any entropy of the system, and is thus
excluded from the proposal.

\begin{figure}[t]
\begin{center}
\includegraphics[width=6in]{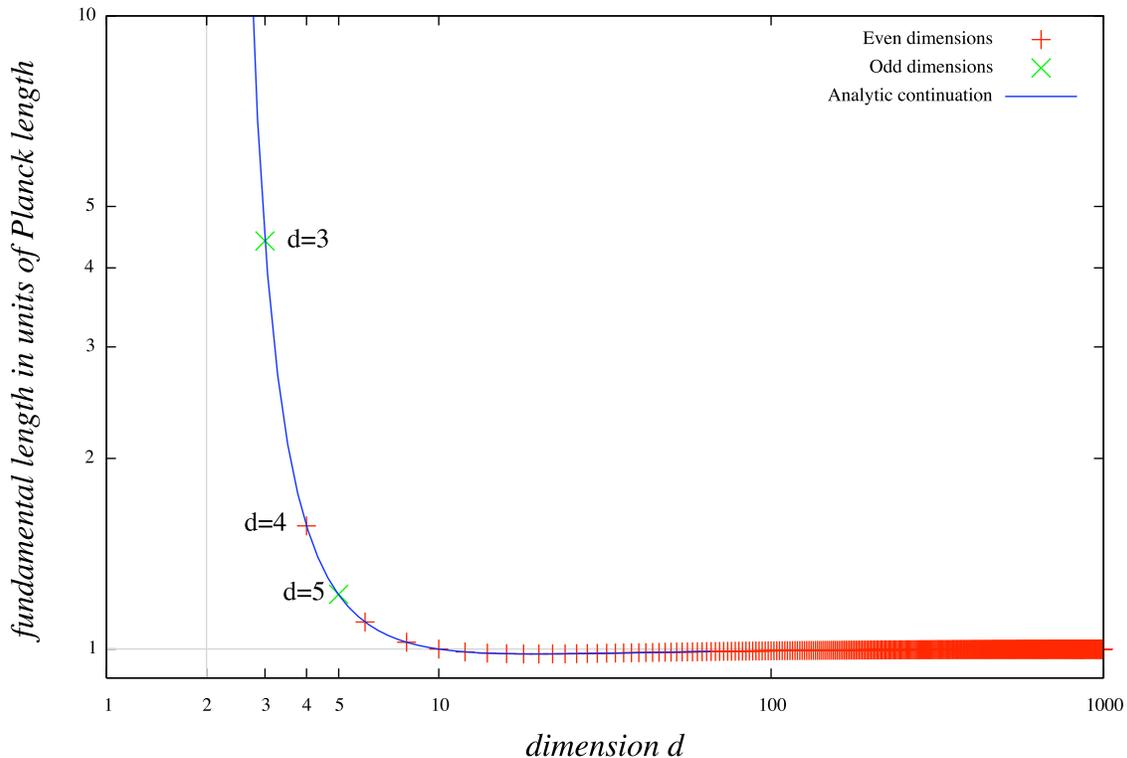}
\caption{Shown is the plot of the fundamental length scale $l_f$ in Planck units as a function of the dimensions $d$ for even and odd dimensions and the analytic continuation.}\label{fig:dimension}
\end{center}
\end{figure}

For large $N$ one can use the asymptotic expansion of the generalized hypergeometric function (App. \ref{app:asymp}) to derive the asymptotics of the number of maximal elements for arbitrary even dimensions, yielding
\begin{equation}
\expec{n}=\frac{2^{\frac{2d-2}{d}}\pi (d-1)}{d\sin\left(\frac{2\pi}{d}\right)\Gamma\left(\frac{2d-2}{d}\right)} \,N^{\frac{d-2}{d}}+...,\quad\text{for}\quad N\rightarrow \infty.
\end{equation} 
From this result one can see that the number of maximal elements scales like $\expec{n}\!\sim\! A/4$, where $A$ is the volume of the boundary of $S_{d-1}(R)$, i.e.
\begin{equation}
A\equiv\Vol\left(\Bcal(S_{d-1}(R))\right)=\frac{\pi^{\frac{d-1}{2}}(d-1)}{\Gamma(\frac{d+1}{2})} R^{d-2}.
\end{equation}
Further, we obtain $\expec{n}\!=\! A/4$ precisely for the following value of the fundamental length scale,
\begin{equation}\label{eq:dimension}
l_f=\left[ \frac{16 \left(\frac{\pi d^2}{4}\Gamma\left(\frac{d+1}{2}\right)^2\right)^{\frac{1}{d}}}{d^2\sin\left(\frac{2\pi}{d}\right)\Gamma\left(2\frac{d-1}{d} \right) }  \right]^{\frac{1}{d-2}}
\end{equation}
The analytic continuation of \eqref{eq:dimension} as a function of the
dimension is shown in Fig. \ref{fig:dimension} together with the explicit
values for 2+1 and 4+1 dimensions as determined in the previous sections. One
observes that the expression \eqref{eq:dimension} agrees with these values.
This suggests that \eqref{eq:dimension} also holds for
arbitrary odd dimensions. For $d=2$ the value of \eqref{eq:dimension}
diverges, since the Planck length $l_p\!=\!(\hbar G_N/c^3)^{(d-2)/2}$ is not
well defined in two dimensions. This also reflects the fact that the entropy
bound only holds for dimensions $d\!\geqslant\!3$. For all values
$d\!\geqslant\!3$ the fundamental length scale is of order of the Planck
length. Interestingly, for large values $d\!\rightarrow\!\infty$ expression
\eqref{eq:dimension} tends exactly to one. It is also less than one for values $d\!>\!d_0$, where $d_0\!\approx\! 10$.

\subsection{Generalizations to different spatial hypersurfaces}\label{sec:hyp}

In the previous section we have derived the Susskind bound for the case where the spacelike hypersurface was chosen to be a $(d-1)$-dimensional ball in $d$ dimensional Minkowski spacetime. Further, from this we determined the fundamental discreteness scale of spacetime.  However, it is important to prove that the fundamental discreteness scale so determined yields the same entropy bound for all spacelike hypersurfaces of a certain dimension. 

\begin{figure}[t]
\begin{center}
\includegraphics[width=4in]{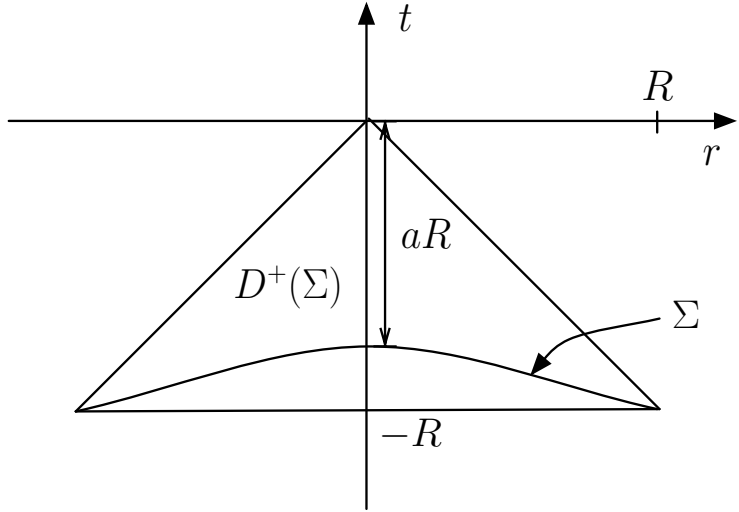}
\caption{Illustration of the different hyperbolic spherically symmetric spatial hypersurfaces $\Scal$ parameterized by $a$, and the domain of dependence $D^+(\Scal)$.}\label{fig:hyp}
\end{center}
\end{figure}

In this section we show that the claim also holds for spacelike
hypersurfaces in Minkowski spacetime different from those discussed in the
previous section. We investigate hyperbolic spherically symmetric spacelike
hypersurfaces given by 
\begin{equation}\label{eq:hypdef}
t = -\sqrt{r^2 + a^2(R^2-r^2)},\quad 0 \leqslant a \leqslant 1,
\end{equation}
as shown in Fig.\ \ref{fig:hyp} together with its domain of dependence. For the special case of $a\!=\! 1$ the spacelike hypersurface given by \eqref{eq:hypdef} is equivalent to the $(d-1)$-dimensional ball $S_{d-1}(R)$ for which we have determined an analytic expression for the expected number of maximal elements $\expec{n}$ in the previous section. For other values $0\!<\!a\!<\!1$ one cannot determine the number of maximal elements analytically. In the following we will investigate this problem numerically for the physically most important cases of 2+1 and 3+1 dimensions.

\subsubsection{2+1 dimensions}

For $d=3$ dimensions we use Monte-Carlo methods to numerically obtain the number of maximal elements in the domain of dependence of the spacelike hypersurfaces $\Scal$ defined by \eqref{eq:hypdef} for different values of $a$ as a function total number $N$ of elements in the domain of dependence. Since all these spacelike hypersurfaces $\Scal$ have the same boundary $\Bcal(\Scal)$, it is useful to express $\expec{n}$ as a function of the length of the boundary $A\!=\! 2\pi R$. One can do this by using that $N\!=\! \rho V$, where the volume of the domain of dependence of $\Scal$ is given by
\begin{equation}
V=\frac{2\pi}{3}\frac{a^2}{1+a}R^3.
\end{equation}
Further, we use the value for the fundamental density of spacetime as obtained
in Sec. \ref{sec:ball2+1}, i.e.\ $\rho\!=\!(3^{5/2}\Gamma(1/3)^3)/(8192 \pi)$.

\begin{figure}[t]
\begin{center}
\includegraphics[width=6in]{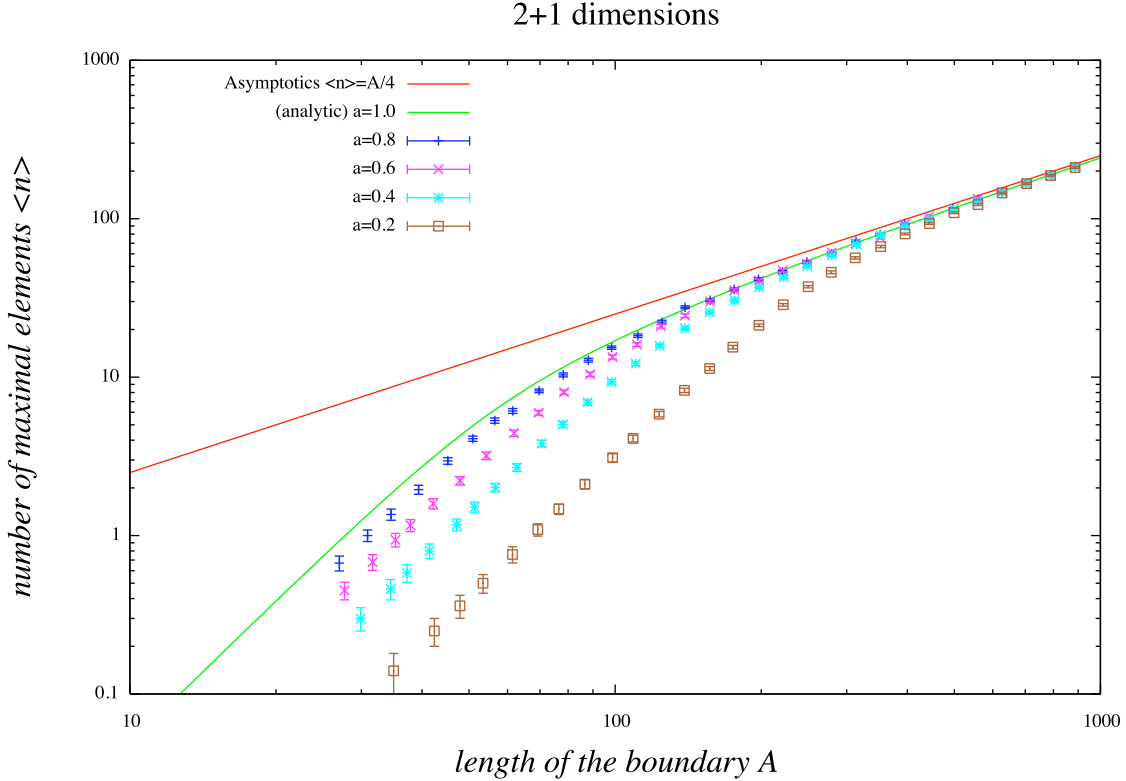}
\caption{Shown is the plot of the expected number of maximal elements $\expec{n}$ in the domain of dependence as a function of the length of the boundary $A$ for different hyperbolic spherical symmetric spacelike hypersurfaces parameterized by $a$ in 2+1 dimensional Minkowski spacetime. One sees that all functions approach the same asymptotic $\expec{n}=A/4$.}\label{fig:hyp2+1}
\end{center}
\end{figure}

The results of the simulations are summarized in Fig.\ \ref{fig:hyp2+1}. Shown
is the expected number of maximal elements $\expec{n}$ as a function of the
boundary length $A$. For the special case of $a\!=\!1$ in \eqref{eq:hypdef} we
know that the result is given by \eqref{eq:nmax2+1}, where $N$ is replaced by
$N\!=\!A^3/(2^{13}\Gamma(2/3)^3)$. The asymptotics of this analytic result is
given by $\expec{n}=A/4$ as shown earlier. For different values of
$0\!<\!a\!<\!1$ the simulations show that even though the number of maximal
elements as a function of $A$ differs for small values $A\!\lesssim\!10^3$ in
Planck units, in the expansion for large $A$, corresponding to the continuum
approximation, all functions $\expec{n}$ for different $a$ enter the same
asymptotic expansion $\expec{n}=A/4$, yielding Susskind's entropy bound. In addition, it shows that
the prediction of the fundamental discreteness scale is universal in 2+1
dimensions, at least for all investigated cases of spacelike
hypersurfaces in Minkowski spacetime. A generalization to examples in curved
spacetime has still to be shown and will be investigated in future work.

\subsubsection{3+1 dimensions}

As in the case of 2+1 dimensions we use Monte-Carlo methods to numerically obtain the number of maximal elements in the domain of dependence of the spacelike hypersurfaces $\Scal$ defined by \eqref{eq:hypdef} for $d=4$ and different values of $a$. Again, the result is expressed as a function of the area of the boundary of $\Scal$, i.e.\ $A\!=\! 4\pi R^2$. This can be done by using the relation $N\!=\! \rho V$ and noticing that the volume of the domain of dependence of $\Scal$ is given by
\begin{equation}
V=R^4 \left[\frac{\pi}{3}-\frac{\pi}{6(1-a^2)^{\frac{3}{2}}}\left(\sqrt{1-a^2}(2-5a^2)+3a^4\log\left( \frac{1+\sqrt{1-a^2}}{a}\right) \right)\right],
\end{equation}
where we use $\rho\!=\!1/6$ for the value of the fundamental density of four dimensional spacetime.

\begin{figure}[t]
\begin{center}
\includegraphics[width=6in]{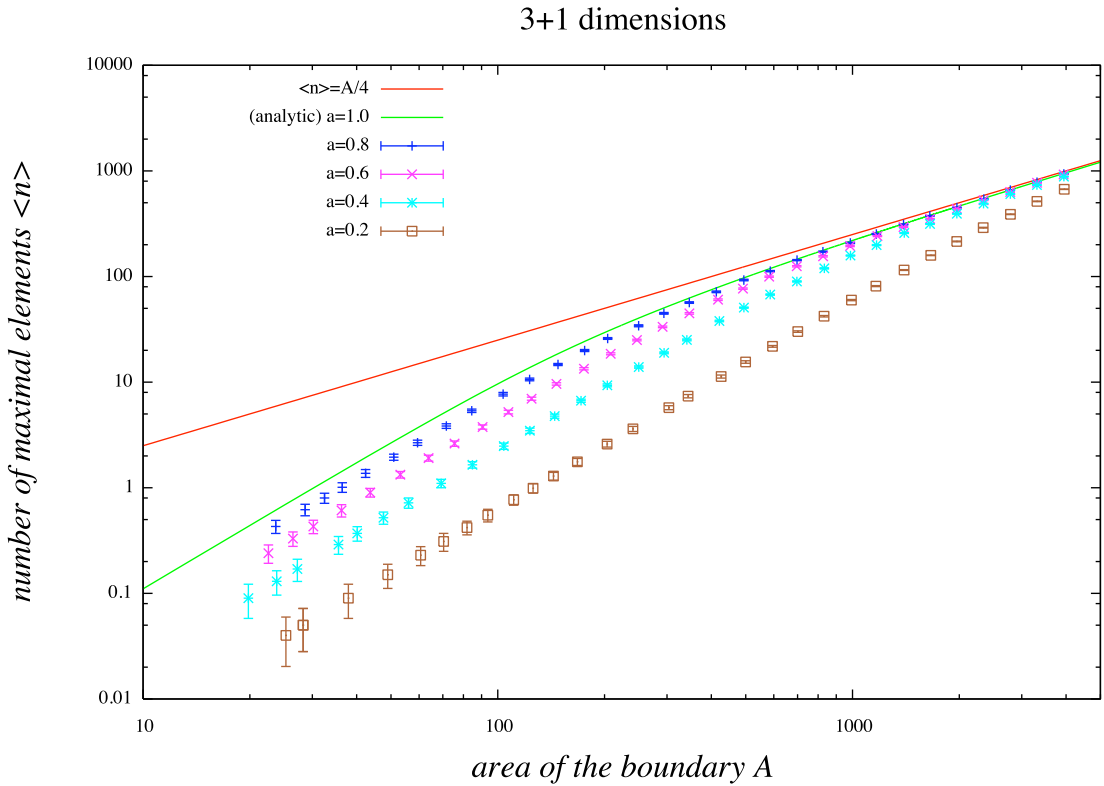}
\caption{Shown is the plot of the expected number of maximal elements $\expec{n}$ as a function of the length of the boundary $A$ for different hyperbolic spherical symmetric spacelike hypersurfaces in 3+1 dimensional Minkowski spacetime parameterized by $a$. One sees that all functions approach the same asymptotics $\expec{n}=A/4$.}\label{fig:hyp3+1}
\end{center}
\end{figure}

The numerical and analytical results are shown in
Fig.\ \ref{fig:hyp3+1}. Displayed is the expected number of maximal elements
$\expec{n}$ as a function of the area $A$ of the boundary. For the special
case of $a\!=\!1$ the analytical result was given by \eqref{eq:nofN3+1}, with
$N$ replaced by $N\!=\!A^2/(2 \pi12^2)$. The simulations show that for
different values of $0\!<\!a\!<\!1$ all functions $\expec{n}$ enter the same
asymptotics, $\expec{n}=A/4$, in the continuum limit, giving further evidence
for the claim. For very small values of $a$ the expected number
of maximal elements enters the asymptotic regime only very slowly. However, the upper
value for $A$ of the simulations is still very small ($\sim\!10^{-33}m$). As
in the lower dimensional case the simulations show that the prediction for the
value of the fundamental discreteness scale of four-dimensional spacetime,
namely $l_f=\sqrt[4]{6}$, is a universal quantity for this dimension, at least
for all investigated cases of spherically symmetric
spacelike hypersurfaces in Minkowski spacetime.

\section{Conclusion}

In this work we motivated causal sets as a candidate for a
theory of fundamentally discrete quantum gravity. Causal structure and discreteness are implemented as fundamental ingredients in this approach and local Lorentz invariance follows as a consequence. In
contrast to attempts to quantize general relativity by quantizing the
metric fields, such as done in loop quantum gravity, causal set theory aims to
formulate quantum gravity as a sum-over-histories, where the single
histories are not smooth Lorentzian manifolds but rather ordered sets of
spacetime elements. Continuum physics is then thought to emerge from this
discrete structure in a continuum approximation.

Using the kinematics of this framework we argued for a bound of the entropy in a spherically symmetric spacelike region from a
counting of potential horizon ``degrees of freedom'' of the fundamental theory, namely
the maximal elements of the future domain of dependence of
the region. It was then shown that, for different spherically
symmetric spacelike regions in Minkowski spacetime of arbitrary
dimension, this leads to Susskind's spherical entropy bound. The evidence was given in terms of analytical results
for spatial $(d-1)$-dimensional balls
in $d$-dimensional Minkowski spacetime, and in terms of numerical results
obtained from Monte-Carlo simulations for the case of hyperbolic spherically
symmetric hypersurfaces in Minkowski spacetime. Agreement with the factor of
proportionality in Susskind's entropy bound required the fundamental discreteness scale
of four-dimensional spacetime to be $l_f=\sqrt[4]{6}\!\simeq\!1.57$ in Planck
units. This is a ``prediction'' for the value of the fundamental
discreteness scale of four dimensional spacetime within the causal set
approach. (The first attempt of a determination of the discreteness scale 
came from computing the entropy of
dimensionally reduced black holes, both Schwarzschild and spherically
symmetric gravitational collapse, as described in \cite{Dou:2003af}.)
Even though this value varied with dimension, it was
shown that, for all investigated cases, 
its value did not depend on the choice of spherically symmetric region for a
given dimension.
However, the evidence provided so far was 
given only in terms
of examples in flat Minkowski spacetime. Clearly this is a 
very restricted class and further evidence should be provided in different spacetimes. Spherically symmetric spacelike regions
in curved spacetime such as in Friedmann-Robertson-Walker cosmology will be
investigated in future work. Another step would be to formulate the proposal
intrinsically in terms of order invariants, without making explicit reference
to the continuum.  Such a formulation 
may be fruitful in providing a fundamental understanding of Bousso's covariant entropy bound.

\section*{Acknowledgments}

We would like to thank R.\ D.\ Sorkin for suggesting the idea of counting
maximal elements as a measure for maximum entropy. Further, we would like to
thank F.\ Dowker for enjoyable discussions, comments, and critical proof
reading of the manuscript. S.Z.\ would also like to thank the theoretical
physics group at Utrecht University for enjoyable discussions on this
work. Funding through the Marie Curie Research and Training Network ENRAGE
(MRTN-CT-2004-005616) is kindly acknowledged.

\appendix

\section{Basic definitions regarding the causal structure}
\label{app:causal}

In this appendix we state some of the basic concepts regarding the causal structure of continuum spacetimes. For further references the reader is referred to \cite{Wald:1984rg,largescale}, where we follow the conventions of \cite{Wald:1984rg}.

Let $(\M,g)$ be a time orientable spacetime. We define a differentiable curve
$\lambda(t)$ to be a \emph{future directed timelike curve} if at each point
$p\elem\lambda$ the tangent $t^{a}$ is a future directed timelike
vector. Further, $\lambda(t)$ is called a \emph{future directed causal curve}
if at each $p\elem\lambda$ the tangent $t^{a}$ is a future directed timelike
or null vector.

Using this definition one can define the chronological future (past) and causal future (past) of a spacetime event or a set of spacetime events.
The \emph{chronological future} of $p\elem\M$, denoted by $I^+(p)$, is defined as
\begin{equation}
\label{eq:cfofp}
I^+(p)=\left\{q\in\M \left|\text{ $\exists$ future directed timelike curve $\lambda(t)$ s.t. $\lambda(0)=p$ and $\lambda(1)=q$} \right.\right\}
\end{equation} 
For any subset $\Scal\subset\M$ we thus define
\begin{equation}
\label{eq:cfofS}
I^+(\Scal)=\bigcup_{p\in\Scal}I^+(p)
\end{equation}
In analogy to $I^+(p)$ and $I^+(\Scal)$ we can also define the \emph{chronological past} $I^-(p)$ and $I^-(\Scal)$ by simply replacing ``future'' by ``past'' in \eqref{eq:cfofp}.

In analogy to the chronological future we can also define the \emph{causal future} of a point $p\elem\M$, denoted by $J^+(p)$, by replacing ``future directed timelike curve'' by ``future directed causal curve'' in \eqref{eq:cfofp}. The definition of $J^+(\Scal)$, $J^-(p)$ and $J^-(\Scal)$ then follow accordingly.

Very important for the formulation of the entropy bound stated in Sec. \ref{sec:conjecture} is the concept of the domain of dependence of a spacetime region.

Let $\Scal$ be any spacelike hypersurface of $\M$.  We define the \emph{future domain of dependence} of $\Scal$, denoted by $D^+(\Scal)$, by
\begin{equation}\label{eq:domain}
D^+(\Scal)=\left\{p\in\M \left|\text{every past inextendible causal curve through $p$ intersects $\Scal$} \right.\right\}.
\end{equation}
Analogously, one can define the\emph{ past domain of dependence} of $\Scal$, denoted by $D^-(\Scal)$, by simply replacing ``past'' by ``future'' in \eqref{eq:domain}.

Using the definitions above one can then define the future and past Cauchy horizon.
Let $\Scal$ be any spacelike hypersurface of $\M$. The \emph{future Cauchy horizon} of $\Scal$, denoted by $H^+(\Scal)$, is defined as
\begin{equation}\label{eq:cauchy}
H^+(\Scal)=\overline{D^+(\Scal)} \setminus I^-(D^+(\Scal)),
\end{equation}
where $\overline{D^+(\Scal)}$ is the closure of  $D^+(\Scal)$. The \emph{past Cauchy horizon}, $H^-(\Scal)$, can be defined in analogy to \eqref{eq:cauchy}. The different concepts defined in this subsection are illustrated in Fig.\ \ref{fig:domain}.

\section{Volume of the causal region between two events}
\label{app:volume}

In this appendix we calculate the volume of the intersection of the causal
future of an event $p$ with the causal past of another event $q\elem J^+(p)$ in Minkowski spacetime $\mathbb{M}^d$, i.e.\ the Alexandrov region $J^+(p)\bigcap J^-(q)$.
This volume depends only on the proper time $\tau$ between $p$ and $q$ and hence we set $\Vol_d(\tau)\!\equiv\!\Vol(J^+(p)\bigcap J^-(q))$. 

Note that the volume of a $d$-dimensional ball with radius $r$, $S_d(r)$, is given by
\begin{equation}
\Vol(S_d(r))=\frac{\pi^{\frac{d}{2}} r^d}{\Gamma(\frac{d}{2}+1)}=:C_d\, r^d, \quad \text{with}\quad C_d=\frac{\pi^{\frac{d}{2}}}{\Gamma(\frac{d}{2}+1)}.
\end{equation}
Using this we obtain
\begin{eqnarray}
\Vol_d(\tau)&=&2\int_0^\frac{\tau}{2}dt\, \Vol(S_{d-1}(t))\nonumber\\
&=&\frac{\pi^{\frac{d-1}{2}}}{2^{d-1} d\, \Gamma(\frac{d+1}{2})}\tau^d=:D_d\, \tau^d,\quad\text{with}\quad D_d=\frac{C_{d-1}}{2^{d-1}d}.
\end{eqnarray}
Specifically we get $\Vol_2(\tau)=\frac{1}{2}\tau^2$, $\Vol_3(\tau)=\frac{\pi}{12}\tau^3$ and $\Vol_4(\tau)=\frac{\pi}{24}\tau^4$.

\section{Asymptotics of the generalized hypergeometric function}
\label{app:asymp}
In this appendix we obtain the asymptotic expansion of the generalized hypergeometric function as given in \eqref{eq:maxgenerald} to first order. 

We use the general expression for the asymptotic expansion of the generalized hypergeometric function ${}_pF_q$ with $p\!=\!q$ which can be derived from \cite{luke}
\begin{eqnarray}
&&\!\!\!\!\!\!\!\!\!\! {}_pF_p(a_1,...,a_p;b_1,...,b_p;z)  =  \left(\prod_{j=1}^p\frac{\Gamma(b_j)}{\Gamma(a_j)}\right) e^z z^\gamma \left(1+\mathcal{O}\left(\frac{1}{z}\right)\right) +\nonumber\\
&+& \left(\prod_{j=1}^p\frac{\Gamma(b_j)}{\Gamma(a_j)}\right) \sum_{k=1}^p \frac{\Gamma(a_k)\prod_{\substack{j=1\\ j\neq k}}^p \Gamma(a_j-a_k)}{\prod_{j=1}^p\Gamma(b_j-a_k)} (-z)^{-a_k} \left(1+\mathcal{O}\left(\frac{1}{z}\right)\right),\quad |z|\rightarrow\infty\label{eq:hypgeomasymp}
\end{eqnarray}
where
\begin{equation}
\gamma=\sum_{k=1}^p a_k -b_k \quad\text{and}\quad a_k\neq a_j \quad\forall\quad k\neq j.
\end{equation}
In the special case of \eqref{eq:maxgenerald} 
we cannot straightforwardly apply \eqref{eq:hypgeomasymp}, since $a_{p-1}\!=\!a_p$. However, since we are only interested in the first order behavior and $a_1\!<\!a_j$ with $a_1\!\neq\!a_j$ for all $ 2\!\leqslant\! j\!\leqslant\! p$, we can obtain the asymptotic expansion of \eqref{eq:maxgenerald} to first order, yielding
\begin{eqnarray}
&&\!\!\!\!\!\!\!\!\!\! {}_{\frac{d}{2}+1}F_{\frac{d}{2}+1}\left( \frac{2}{d},\frac{4}{d},...,\frac{d}{d},1; 
\frac{d+1}{d},\frac{d+3}{d},...,\frac{2d-1}{d},2,-2^{1-d}N
\right) \nonumber  \\
& = &\left(\prod_{j=1}^{d/2+1} \frac{\Gamma(b_j)}{\Gamma(a_j)}\right)  \frac{\Gamma(a_1)\prod_{j=2}^{d/2+1} \Gamma(a_j-a_1)}{\prod_{j=1}^{d/2+1}\Gamma(b_j-a_1)} \left(2^{1-d}N\right)^{-a_1}+ \mathcal{O}(N^{-2a_1}), \quad N\rightarrow \infty \nonumber\\
&=& \frac{d-1}{d}\frac{2^{\frac{2d-2}{d}} \pi}{\sin\left(\frac{2\pi}{d}\right)\Gamma\left(\frac{2d-2}{d}\right)} N^{-\frac{2}{d}}+ \mathcal{O}(N^{-\frac{4}{d}}), \quad N\rightarrow \infty.
\end{eqnarray}

\addcontentsline{toc}{section}{References}
\providecommand{\href}[2]{#2}\begingroup\raggedright\endgroup

\end{document}